\documentclass[%
 reprint,
superscriptaddress,
bibnotes,
 amsmath,amssymb,
 aps,
prb,
longbibliography
]{revtex4-2}

\usepackage{graphicx}
\usepackage{dcolumn}
\usepackage{bm}

\usepackage[colorlinks=true,citecolor=blue,linkcolor=blue, urlcolor=cyan]{hyperref} 

\usepackage{empheq}
\usepackage{xcolor}
\usepackage[normalem]{ulem}
\usepackage[english]{babel}

\begin{document}


\title{Electromagnetic field assisted exciton diffusion in moir\'e superlattices}

\author{A.M. Shentsev}
\affiliation{Moscow Institute of Physics and Technology, Dolgoprudny, Russia}
\affiliation{L. D. Landau Institute for Theoretical Physics, 142432 Chernogolovka, Russia}
\author{M.M. Glazov}
\affiliation{Ioffe Institute, 194021 St. Petersburg, Russia }%

\date{\today}

\begin{abstract} 
We study exciton energy spectrum and their propagation in moir\'e superlattices formed in transition metal dichalcogenide heterobilayers. In such structures, as a result of weak interlayer interaction, an effective, moir\'e, potential acting on excitons arises. Usually, excitons are considered to be localized in such potential. Here we demonstrate that the coupling of optically active excitons with induced electromagnetic field produces linear in the wavevector energy dispersion even if the quantum mechanical tunneling between the localization sites is suppressed. The effect can be described as a result of the processes of virtual generation-recombination of excitons at the localization sites that results in the $ r^{-3}$ dependence of the transfer matrix element on the intersite distance $r$. Based on the calculated energy spectrum we study exciton propagation in moir\'e superlattices with allowance for the light-exciton interaction. We consider semiclassical diffusion of excitons and take into account exciton-phonon and exciton-static defect scattering. For these mechanisms the diffusion coefficient decreases with increase of the temperature. We also 
analyze the hopping propagation regime and demonstrate that the temperature dependence of the exciton diffusion coefficient is described by the power-law rather than by an exponential function of the temperature. 
\end{abstract}

\maketitle

\section{\label{sec:intro}Introduction}
Heterostructures composed of van der Waals materials possess spectacular physical properties and attract wide scientific interest~\cite{Geim:2013aa}. Lack of strong chemical bonding between the layers gives rise to novel degrees of freedom in such systems, particularly, twist of the layers, absent in conventional semiconductors. It gives rise to novel physical effects, including built-in chirality~\cite{PhysRevB.105.L241406} and emergence of moir\'e patterns, new superstructures absent in the basic materials. They arise due to the small relative mismatch of lattice constants and the rotation between the layers. In gapless graphene, the moiré effect results in formation of flat bands~\cite{bistritzer2011moire} providing a playground for studying strongly correlated electronic systems~\cite{Hunt:2013kx}. All-in-all, moir\'e systems attract an increasing interest~\cite{Xiao_2020,kennes2021moire,Morales-Duran:2024aa}.

Among various van der Waals systems actively studied nowadays, also with prospects for moir\'e effects, heterostructures based on transition metal dichalcogenides take a special place. Transition metal dichalcogenide monolayers are  two-dimensional semiconductors with a direct band gap of $\sim 2$ eV with unusual band structure and optical properties determined by excitons, bound states of electron and hole~\cite{RevModPhys.90.021001,Durnev:2018,Datta:2024aa}. Large binding energies of excitons and other Coulomb-bound complexes open up many prospects for studying interacting systems in two-dimensions with potential applications~\cite{Glazov:2024aa}.

In transition metal dichalcogenide-based van der Waals heterostructures, moir\'e effect leads to modulation of the band gap, which creates an additional potential acting on excitons potentially resulting in states with non-trivial topology~\cite{wu2017topological}. This, together with the importance of excitonic states for two-dimensional semiconductors had ignited active experimental and theoretical studies of excitonic effects in transition metal dichalcogenide heterobilayers where the moir\'e effects are expected to be particularly prominent~\cite{Tartakovskii:2020aa}.

So far, most of the studies focused on the energy spectrum and optical properties of excitons in moir\'e heterostructures~\cite{alexeev2019resonantly,jin2019observation,seyler2019signatures,tran2019evidence,Baek:2021aa,PhysRevX.11.031033,Campbell:2022aa} and on application of excitons to probe manybody correlations of electrons~\cite{Xu:2020aa,Ciorciaro:2023aa,Tan:2023aa}. It is because relatively large potential barriers formed in moir\'e superlattices prevent efficient exciton tunneling and the resulting energy minibands are quite narrow. On the other hand, transport properties of excitons are now highly topical~\cite{Chernikov:2023ab}, particularly, in heterostructures~\cite{Malic:2023aa,Tagarelli:2023aa,PhysRevLett.132.016202,D4NR00136B}. Pioneering works aimed at studies of exciton transport in moir\'e heterostructures have revealed slow exciton propagation~\cite{li2021interlayer}.

Here we demonstrate that excitons can be delocalized and efficiently propagate in moir\'e superlattices based on transition metal dichalcogenide monolayers owing to their interaction with electromagnetic field. We show that the coupling of the excitons with produced electromagnetic field results in the formation of the propagating, longitudinal, excitonic branch with linear dispersion despite the fact that in the absence of the light-matter coupling the excitons are localized at the potential minima, cf. Refs.~\cite{Baimuratov:2013aa,PhysRevLett.128.217402,Li_2024}. The origin of the effect is the long-range exchange interaction between an electron and a hole in exciton~\cite{BP_exch71,PhysRevB.41.7536,Yu:2014fk-1,glazov2014exciton,Iakovlev:2024aa,glazov2024ultrafastexcitontransportvan} that results in virtual recombination of the exciton on one moir\'e lattice site and its formation at another one similarly to the F\"orster process~\cite{Forster:1948aa,agranovich:galanin,Baimuratov:2020aa}. It can be also interpreted as a result of the dipole-dipole interaction in the system. We propose a simple model of the light-matter interaction of excitons considering them as point dipoles localized at the sites of a moiré superlattice. The energy spectrum and eigenmodes of the system are calculated in Sec.~\ref{sec:ddinter}. Based on these results we study in Sec.~\ref{sec:transport} the exciton propagation and evaluate the diffusion coefficient in the semiclassical regime where excitons weakly scatter off the static disorder and phonons (Sec.~\ref{sec:semicldiff}) and in the hopping regime realized for sufficiently strong disorder (Sec.~\ref{sec:hopping}). We predict a power-law temperature dependence of the exciton diffusion coefficient in moir\'e superlattices both in semiclassical and hopping regimes.

\section{\label{sec:ddinter}Dipole-dipole interaction and exciton spectra}
\subsection{Exciton spectrum}
\begin{figure*}[t]
    \centering
    \includegraphics[width=\textwidth]{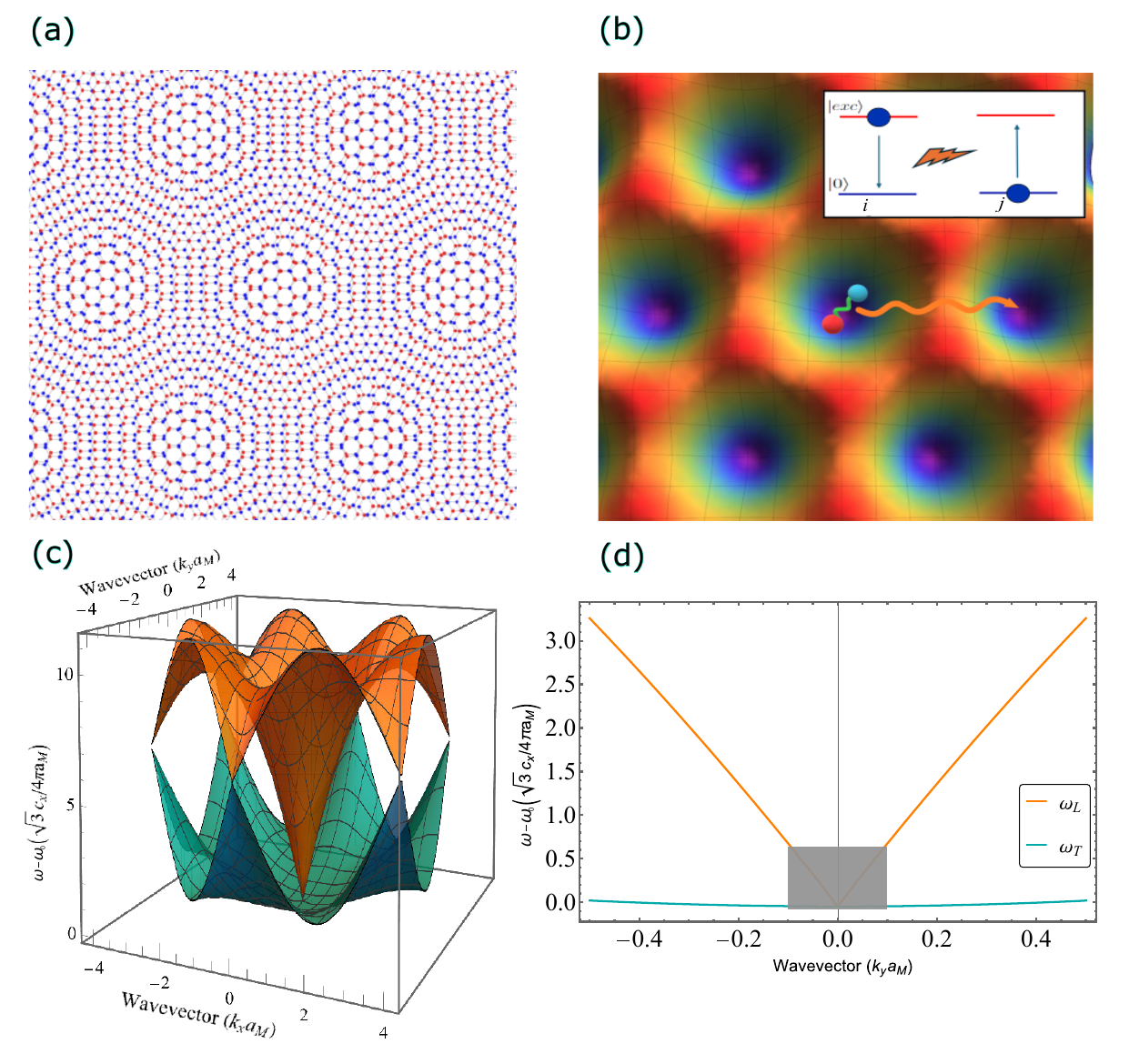}
    \caption{(a) Schematic illustration of the moir\'e pattern formed in twisted heterobilayer. Red and blue dots show the transition metal and chalcogen atoms, respectively. The twist angle $\theta=5^\circ$, lattice constants mismatch is neglected. (b) False color plot of the model potential experienced by the exciton. Inset shows mechanism of exciton propagation between the sites $i$ and $j$ assisted by virtual recombination-generation processes. (c) Exciton dispersion calculated by Eqs.~\eqref{dipole:latt:bloch} (d) Exciton dispersion in the vicinity of the $\Gamma$-point. Shaded area shows the part of dispersion near a light cone, at $k<\omega_0/c$ excitons are radiatively damped.}
    \label{fig1}
\end{figure*}

The system under study is illustrated in Fig.~\ref{fig1} where panel (a) shows a formation of moir\'e lattice  in a twisted heterobilayer and panel (b) depicts the effective potential experienced by the excitons. {We consider a heterobilayer like, e.g., MoSe$_2$/WSe$_2$ or MoSe$_2$/WS$_2$. The specifics of the material composition mainly affects the band alignment,  parameters of moir\'e potential, exciton transition energy, and its oscillator strength, but does not affect general conclusions of the work.} Owing to  weak interlayer interaction, a modulation of the band gap takes place, which leads to the appearance of an effective periodic potential acting on excitons~\cite{wu2017topological}. The potential minima form a triangular lattice~\cite{bistritzer2011moire,PhysRevB.89.205414}, see Fig.~\ref{fig1}(a,b). The model of ideal periodic lattice~\cite{PhysRevB.89.205414} is a reasonable idealization: we neglect weak aperiodicity effects and possible mesoscopic reconstruction~\cite{Zhao:2023aa}. Therefore, our calculations are performed for this geometry, although the results obtained are qualitatively independent of the lattice type.

Estimates show that for realistic depths of potential wells of approximately $\sim 10 \ldots 100 $ meV and with a moiré period of $a_M{\gtrsim} 10$ nm, quantum mechanical tunneling of excitons with {the total mass} $\sim m_{0}$, where $m_{0}$ is the free electron mass, is negligible~\cite{note:tun}. However, optically active -- bright -- excitons produce electromagnetic field that can propagate and excite exciton at a different lattice site. To describe this effect we consider a radiative doublet of excitonic states polarized in the monolayer plane (the specific selection rules depend on the stacking~\cite{Yu:2018aa,Forg:2019aa}). The exciton at a given lattice site $i$ produces a microscopic dipole moment $\bm d_i=(d_{x,i},d_{y,i})$ being a quantum mechanical average of the dipole moment operator 
\begin{equation}
\label{d:op}
\hat{d}_{\alpha,i} = d_{exc}\hat{a}^\dag_{\alpha,i} + {\rm h.c.},
\end{equation}
where $\alpha=x,y$ denotes the polarization, $d_{exc}$ is the dipole moment of per exciton that is determined by the interband momentum matrix element and envelope function, see below and Refs.~\cite{ivchenko05a,PhysRevA.109.053523}, $\hat{a}^\dag_{\alpha,i}$ is the creation operator of an $\alpha$-polarized exciton at the site $i$, and ${\rm h.c.}$ stands for the Hermitian conjugate. We assume that the excitons are described as point dipoles, this approximation is justified since we are interested in the coupling between the remote lattice sites separated by the distances that exceed by far both the moir\'{e} lattice period $a_M$, exciton localization radius $a_{loc}$, and exciton radius $a_{exc}$. It is the coupling between the remote lattice sites that determines exciton dispersion in the relevant, small wavevector range, we shown below.

The dipole moments $\bm d_i$ oscillate with time at a frequency $\omega$ close to $\omega_0$, the exciton responance frequency. The interaction between the lattice sites and, correspondingly, the exciton transfer occurs via the electric field induced by the excitons. The field at the site $i$ can be written as
\begin{equation}
\label{field:greens}
    \hat{\bm{E}}_i = 4\pi q^2\sum_{j \neq i}G_{\alpha\beta}(\bm r_i - \bm r_j)\hat{\bm{d}}_{\beta,j},
\end{equation}
where $\bm r_i$ and $\bm r_j$ are the coordinates of the corresponding lattice sites, $G_{\alpha\beta}(\bm r)$ is the dyadic Green's function (Green's function of Maxwell's equations)~\cite{ivchenko05a,PhysRevA.109.053523}:
\begin{multline}
\label{Greens}
 G_{\alpha\beta}(\bm r)=-\frac{\delta_{\alpha\beta}}{3q^2}\delta(r)+\\ \frac{e^{\mathrm i qr}}{4\pi r}
 \left[\frac{2}{3}\delta_{\alpha\beta}+\left(\frac{3r_\alpha r_\beta}{r^2}-\delta_{\alpha\beta}\right)\left(\frac{1}{(qr)^2}-\frac{\mathrm i}{qr}-\frac {1}{3}\right)\right],
\end{multline}
$q=\omega/c$ is the wavevector of radiation at a frequency $\omega$ (to be self-consistently found, see below), and the summation over the repeated subsripts is implicitly assumed. For simplicity we consider a moir\'e lattice a free space, one can take into account the dielectric environment by choosing an appropriate Green's function.
Taking into account that the exciton interaction with the field is given by the Hamiltonian $-\hat{\bm d} \cdot \hat{\bm E}$ we arrive at the exciton Hamiltonian in the following form:
\begin{equation}
\label{Ham}
    \hat{\mathcal H} = \hat{\mathcal H}_0 - 2\pi q^2\sum_{i,j\ne i}\hat{\bm{d}}_i\hat{G}(\bm r_i - \bm r_j)\hat{\bm{d}}_{j},
\end{equation}
where the summation goes over all lattice sites $i$ and $j$ [the term with $i=j$ is excluded to avoid diverging self-interaction and the difference in prefactors in Eqs.~\eqref{field:greens} and \eqref{Ham} is to avoid double counting of the pairs of sites in Eq.~\eqref{Ham}],
\begin{equation}
\label{Ham:0}
\hat{\mathcal H}_0 = \sum_{i,\alpha} \hbar\omega_0 \hat{a}^{\dagger}_{\alpha,i}\hat{a}_{\alpha,i}
\end{equation}
is the Hamiltonian of bare excitons in the absence of tunneling and coupling with electromagnetic field, we recall that $\omega_0$ is the exciton resonance frequency.

Quantum mechanical Hamiltonian~\eqref{Ham} describes a lattice of coupled dipoles. It is convenient to consider corresponding classical problem presenting the equation for the dipole moment $\bm d_i \propto \exp{(-\mathrm i \omega t)}$ as
\begin{equation}
\label{p:latt}
     \hbar(\omega_0 -\omega)\bm{d}_i= 4\pi q^2|{d}_{exc}|^2\sum_{j \neq i} \hat{G}(\bm r_i - \bm r_j)\bm{d}_j.
\end{equation}
{The derivation of Eqs.~\eqref{p:latt} without point-dipole assumptions and the analysis of excitonic states is presented in Appendix~\ref{sec:append}.}

According to the Bloch theorem, we will look for solutions in the form:
\begin{equation}
\label{dipole:bloch}
    \bm{d}_{i} = e^{-\mathrm i\omega t + \mathrm i \bm{k}\cdot\bm{r}_{i}}\bm{d}_{\bm{k}},
\end{equation}
where $\bm k$ is the quasi wavevector. For $k \gg q$ we can neglect the retardation in light propagation between the lattice sites in Eq.~(\ref{Greens}) and take into account the leading terms only. As a result, we obtain
\begin{equation}
\label{dipole:latt:bloch}
     \hbar(\omega_0 -\omega)d_{\alpha,\bm{k}}= |{d}_{exc}|^2\sum_{\bm{r} \neq 0} e^{\mathrm i\bm{k}\bm{r}}\frac{3r_{\alpha}r_{\beta} - r^2\delta_{\alpha\beta}}{r^5}d_{\beta,\bm{k}},
\end{equation}
where the origin of coordinates $\bm r = 0$ is taken at one of the superlattice sites and summation is taken over the lattice sites. For a particular triangular lattice depicted in Fig.~\ref{fig1}(a,b) the coordinates of the lattice sites are given by
\begin{equation}
 \bm r_{i}= a_M\left(\frac{1}{2}n+m, \frac{\sqrt{3}}{2}n \right), \quad i \equiv (n,m). 
\end{equation}
The lattice sites $i$ are enumerated by a pair of integers $n,m \in \mathbb Z$. Equation~\eqref{dipole:latt:bloch} describes the dipole-dipole interaction between the lattice sites.

The exciton dispersion in moir\'e superlattice calculated from solution of Eq.~\eqref{dipole:latt:bloch} is plotted in Fig.~\ref{fig1}(c)
 across the whole mini Brillouin zone and in Fig.~\ref{fig1}(d) the range of relatively small wavevectors $ka_M \ll 1$. Two branches in dispersion are clearly seen: for one of the branches (bottom, denoted as `T') the dependence $\omega(\bm k)$ is almost absent, while for the other one (top, denoted as `L') the $\omega(\bm k)$ grows linearly with the wavevector. 
 
 In the vicinity of the $\Gamma$-point, $k a_M \ll 1$, one can pass to a continuous limit in Eq.~\eqref{p:latt}. To do this, we represent the Green's function as:
 \begin{equation}\label{eq:Gfree:1}
 G_{\alpha\beta}(\bm r)=\left(\delta_{\alpha\beta} + \frac{1}{q^2} \frac{\partial^2}{\partial r_\alpha \partial r_\beta} \right) \frac{e^{\mathrm i qr}}{4\pi r}.
\end{equation}
Evaluating the Fourier transform of the Green's function~\eqref{eq:Gfree:1} taking into account that all $\bm r_i$ are in the same plane we have instead of Eq.~\eqref{p:latt}
\begin{multline}
\label{p:cont}
    (\omega_{0}-\omega)d_{\alpha,\bm{k}} =\\ \frac{c_x q^2}{\sqrt{k^2 - q^2}}\sum_\beta \left(\delta_{\alpha\beta} - \frac{k_{\alpha}k_{\beta}}{q^2}\right)d_{\beta,\bm{k}},
\end{multline}
where
\begin{equation}
    \label{cx}
    c_{x} = \frac{4\pi|d_{exc}|^2}{\sqrt{3}\hbar a_M^2}.
\end{equation}
At $\bm k\to 0$, but for the states outside of the light cone, $k\gg q$, we obtain 
\begin{subequations}
\label{dispersion}
\begin{equation}
\label{dispersion:L}
    \omega_L(\bm k) =  \omega_{0} + c_x k, 
\end{equation}
and
\begin{equation}
\label{dispersion:T}
    \omega_T(\bm k) \approx \omega_{0}.
\end{equation}
\end{subequations}
The analysis of the eigenvectors shows that indeed the branch `L' corresponds to a longitudinal exciton, i.e., the state where $\bm d_{\bm k} \parallel \bm k$, while the branch `T' corresponds to a transversal exciton, where $\bm d_{\bm k}\perp \bm k$. In agreement with the general analysis and previous works~\cite{Baimuratov:2013aa,PhysRevLett.128.217402,Li_2024,BP_exch71,PhysRevB.41.7536,Yu:2014fk-1,glazov2014exciton,Iakovlev:2024aa,glazov2024ultrafastexcitontransportvan} the longitudinal branch becomes dispersive. Such excitons propagate with the speed $c_x$ despite the absence of quantum mechanical tunneling. The regime $k\gg q$ corresponds to the situation where the retardation of electromagnetic field propagation can be neglected and exciton couplings occurs due to curlless `near' field. This regime corresponds to the electrostatic $1/r^3$ asymptotic of the Green's function~\eqref{Greens}. In such a case the effect can be interpreted as a result of the dipole-dipole interaction. Its long-range nature results in a non-analytic $\propto k$ contribution to the exciton dispersion.





\subsection{Estimation of group velocity of excitons}\label{subsec:est}
Let us now estimate the propagation velocity of excitons in moir\'e superlattice. It follows from Eq.~\eqref{cx} and \eqref{dispersion:L} that the velocity $c_x$ is determined by the superlattice period and the microscopic dipole moment of exciton, i.e., by the light-matter interaction parameters, while the height of the barriers and depth of potential wells do not affect $c_x$ explicity. The exciton dipole moment is estimated as follows~\cite{ivchenko05a,PhysRevA.109.053523}
\begin{equation}
\label{d:exc:est}
    d_{exc} = d_{cv}\int d\bm \rho \psi(\bm \rho, \bm \rho) \sim d_{cv}\frac{a_{loc}}{a_{exc}},
\end{equation}
where $d_{cv}$ is the interband dipole matrix element and $\psi(\bm \rho_e, \bm \rho_h)$ is the envelope function of the in-plane motion of electron and hole taken at the coinciding coordinates of the charge carriers. The last estimate in Eq.~\eqref{d:exc:est} is performed assuming that the exciton is localized as a whole such that $\psi(\bm \rho, \bm \rho)$ can be recast as a product of the relative motion envelope function with effective Bohr radius $a_{exc}$ and the center-of-mass envelope with effective localization radius $a_{loc}$, see Ref.~\cite{Semina_2022} {and Appendix~\ref{sec:append}} for details. Expressing $d_{cv}$ via the interband momentum matrix element $p_{cv}$ and making use of the value $\gamma = \hbar p_{cv}/m \sim 2-3 ~\text{eV}$~\AA~\cite{kormanyos2015k} we have for bright intralayer excitons
\begin{equation}
\label{estimate:c_x}
    c_x \sim  \frac{e^2}{\hbar}\left(\frac{ \gamma}{\hbar\omega_0}\right)^2\times\frac{4\pi}{\sqrt{3}a^2_{M}}\sim 4\cdot 10^{4} ~\text{m}/\text{s},
\end{equation}
where we took $\omega_0 \sim 10^{15}~\text{s}^{-1}$ ($\hbar\omega_0 = 1$~eV), $a_{M} \sim 5 ~\text{nm}$ and for a rough estimate we assumed $a_{exc} \sim a_{loc}$. For interlayer excitons the $d_{cv}$ is smaller since it contains small tunneling parameter and, correspondingly, for interlayer excitons the velocity $c_x$ is smaller than the estimate~\eqref{estimate:c_x}. Depending on the system parameters, the speed of longitudinal excitons is $\sim 0.01-0.1\%$ the speed of light in vacuum.

{Above, a moir\'e heterostructre freely suspended in vacuum has been studied. An encapsulation results in reduction of exciton propagation velocity because of the screening of the electromagnetic field similarly to the screening of the long-range exchange interaction in excitons~\cite{prazdnichnykh2020control}. For a moir\'e superlattice surrounded by a homogeneous dielectric media with the dielectric permittivity $\varepsilon_b(\omega)$ the propagation velocity is reduced by a factor of $\varepsilon_b(\omega_0)$ compared to the value in Eq.~\eqref{cx}:
\[
    c_{x} = \frac{4\pi|d_{exc}|^2}{\varepsilon_b(\omega_0)\sqrt{3}\hbar a_M^2},
\]
while in multilayered van der Waals heterostructures the $\varepsilon_b(\omega_0)$ should be replaced by an effective quantity which weakly depends on the exciton wavevector and on the system's composition and geometry~\cite{prazdnichnykh2020control}.}

\section{Transport of longitudinal excitons}\label{sec:transport}
Now we turn to the problem of exciton propagation in moir\'e lattices. As shown above, the longitudinal exciton has a dispersion in the absence of tunneling between the lattice sites that appears as a result of its coupling with electromagnetic field. In an ideal, defect-less system the excitons would propagate ballistically with the speed $c_x$. In real structures there are inevitable static defects related to the impurities and imperfections of moir\'e pattern. At finite temperatures excitons interact with phonons. As a result, free  propagation is interrupted by scattering and excitons move diffusively rather than ballistically. We are interested in the diffusion coefficient $D$ of excitons, which relates the flux density of quasiparticles $i$ and their concentration or density gradient $n$~\cite{Chernikov:2023ab}
\begin{equation}
    \bm i = -D\bm\nabla n.
\end{equation}
In what follows we consider two basic regimes of exciton transport: (i) semiclassical propagation where the scattering is relatively weak and can be described within the Boltzmann kinetic equation theory and (ii) hopping regime that is realized for strong disorder where the excitons are mainly locaized at the lattice sites and rarely jump between them with emission or absoprtion of phonons.

\subsection{Semiclassical regime}\label{sec:semicldiff}

The semiclassical propagation of excitons is realized if the temperature is sufficiently high such that 
\begin{equation}
    \label{condition:Boltzmann}
    k_B T \langle\tau\rangle_\varepsilon \gg \hbar,
\end{equation}
where $\langle\tau\rangle_\varepsilon $ is the effective momentum relaxation time. It can be expressed as a thermal average 
\begin{equation}
    \label{tau:aver}
     \langle\tau\rangle_{\varepsilon} = \frac{\int  \tau(\varepsilon)e^{-\varepsilon/k_BT}\nu(\varepsilon)d\varepsilon}{\int  e^{-\varepsilon/k_BT}\nu(\varepsilon)d\varepsilon},
\end{equation}
where $\nu(\varepsilon) \propto \varepsilon$ is the energy-dependent density of states of longitudinal excitons, $\varepsilon\equiv \varepsilon(\bm k) =  \hbar\omega_L(\bm k)$, and $\tau(\varepsilon)$ is the energy-dependent momentum relaxation time introduced below for the point defects and long-wavelength acoustic phonon scattering. The exciton diffusion coefficient in the $\tau-$approximation for the elastic or quasi-elastic scattering can be expressed as~\cite{Chernikov:2023ab,Glazov:2022aa}
\begin{equation}
\label{eq:D:tau}
    D_{cl} = 
    \left\langle\frac{v_k^2}{2}\tau(\varepsilon) \right\rangle
    = \frac{c_x^2}{2}\langle\tau\rangle_{\varepsilon},
\end{equation}
with $\bm v_{\bm k} = \partial \omega_L/\partial \bm k$. Note that under the condition~\eqref{condition:Boltzmann} the splitting of the longitudinal and transverse branches of excitons is large,  for relevant wavevectors $\omega_L(\bm k) - \omega_T(\bm k) \gg \langle\tau\rangle_\varepsilon^{-1}$, [cf. Eqs.~\eqref{dispersion}]. Thus, the longitudinal and transverse excitons can be considered independently. Equation~\eqref{eq:D:tau} provides the diffusion coefficient for the longitudinal excitons, for the transverse excitons $D=0$ since they are dispersionless and their group velocity vanishes. Below we analyze Eq.~\eqref{eq:D:tau} for two main scattering mechanisms.

\subsubsection{Scattering by point defects}
Let us consider exciton scattering by short-range defects using perturbation theory. We write the perturbation potential induced by a point defect at the position $\bm R_i$ as the change in exciton energy at the site as a $2\times 2$ matrix acting in the basis of two excitonic states polarized along the $x$ and $y$ axes of the crystal:
\begin{equation}
\label{x:def}
    \hat{\mathcal U}_i  = \begin{pmatrix}
        1 & 0\\
        0 & 1
    \end{pmatrix}u \, \delta(\bm r - \bm R_i).
\end{equation}
We assume that the defect does not break the symmetry of the lattice and all defects are identical. It follows then that the matrix element of the longitudinal exciton scattering can be written as
\begin{equation}
    U_{\bm k', \bm k} = \frac{u\cos{\varphi}}{S}\sum_{i}e^{-\mathrm i(\bm{k'}-\bm{k})\bm{R}_i}.
\end{equation}
Here the summation over all defects is assumed, $\varphi$ is the scattering angle, i.e., the angle between the initial, $\bm k$, and final, $\bm k'$, wavevectors of the excitons, $S$ is the (normalization) crystal area. The factor $\cos{\varphi}$ results from the fact that we consider longitudinal excitons only. Neglecting an interference at scattering off different defects we get
\begin{equation}
\label{eq:V:res}
    |U_{\bm k', \bm k}|^2 \approx \frac{nu^2}{S}\cos{^2\varphi},
\end{equation}
where $n$ is the density of defects. Making use of  Fermi's golden rule 
\begin{equation}
    \label{tau:FGR}
    \frac{1}{\tau_{def}(\varepsilon)} = \frac{2\pi}{\hbar} \sum_{\bm k'}   |U_{\bm k', \bm k}|^2(1- \cos{\varphi}) \delta[\varepsilon(\bm k') - \varepsilon(\bm k)],
\end{equation}
we have
\begin{equation}
\label{tau:def:fin}
    \frac{1}{\tau_{def}(\varepsilon)} = \varepsilon\frac{nu^2}{2\hbar^{3}c_x^{2}}.
\end{equation}
Hence, for the diffusion coefficient caused by the scattering by static defects, we obtain from general Eq.~\eqref{eq:D:tau} the following expression:
\begin{equation}
\label{Diffus:def}
    D_{cl}^{def} 
    = \frac{2}{k_BT}\frac{\hbar^3c_x^4}{2nu^2} \propto T^{-1}.
\end{equation}
Note that an increase in the temperature $T$ results in the reduction of the diffusion coefficient. This is because the scattering rate~\eqref{tau:def:fin} grows linearly with increasing the exciton energy and, hence, the temperature owing to the linear-in-$\varepsilon$ density of states, while exciton velocity is energy independent.
Let us emphasize the difference between this result and the diffusion coefficient of excitons with parabolic dispersion, where $D_{cl} \propto T$ for scattering on static “short-range” defects. In the latter case the scattering rate is energy independent while the $v_k^2$ is a linear function a temperature~\cite{Glazov:2022aa}.

\subsubsection{Scattering by longitudinal acoustic phonons}
Phonons serve as an important omnipresent source of scattering. For encapsulated transition metal dichaclogenides long wavelength acoustic phonons provide the key contribution to the scattering controlling the exciton linewidth and diffusion~\cite{C9NR04211C,Chernikov:2023ab}. Accordingly, let us consider this basic mechanism of scattering of excitons by the acoustic phonons, described by the deformation potential interaction:
\begin{multline}
\label{eq:defpot}
    \hat{\mathcal U}^{DP} = \mathrm i \Xi\begin{pmatrix}
        1 & 0\\
        0 & 1
    \end{pmatrix}\times\sum_{\bm q} q\left(\frac{\hbar}{2\rho \omega_{LA}(\bm{q})S}\right)^{1/2}\times\\(\hat{b}_{LA,q}e^{\mathrm i\bm q\bm r-\mathrm i\omega t} - \hat{b}^{\dagger}_{LA,q}e^{-\mathrm i\bm q\bm r+\mathrm i\omega t}).
\end{multline}
Here $\omega_{LA}(\bm q) = sq$ with $s$ being the speed of sound is the dispersion of longitudinal acoustic phonons in the relevant range of small wavevectors $\bm q$, $\Xi$ is the deformation potential constant being a combination of the deformation potentials for the electron and the hole~\cite{shree2018observation}, $\rho$ is the two-dimensional mass density of the crystal. Similarly to the exciton-defect scattering, Eq.~\eqref{x:def}, the perturbation is a $2\times 2$ matrix in the space of $x$- and $y$-polarized excitons.
Estimates according to Eq.~\eqref{estimate:c_x} and data from~\cite{shree2018observation} show that $c_x/s \sim 10-100$, i.e., phonons are considerable slower than excitons. This means that a phonon with the same momentum has a significantly lower energy compared to an exciton, so the exciton-phonon scattering is quasi-elastic. It allows one to simplify the collision integral, ignoring the difference in the rates of emission and absorption of phonons and neglecting the phonon energy in the energy conservation law. As a result, the Fermi's golden rule takes the form
\begin{equation}
    \label{tau:FGRSP}
    \frac{1}{\tau_{ph}(\varepsilon)} = \frac{2\pi}{\hbar} \sum_{\bm k'}   |M_{\bm k', \bm k}|^2(1- \cos{\varphi})N(q) \delta[\varepsilon(\bm k') - \varepsilon(\bm k)],
\end{equation}
where $N(\bm q)$ is the phonon occupancy and $M_{\bm k \bm k'}$ is the matrix element of the longitudinal exciton-phonon interaction:
\begin{equation}
\label{matr:el:ph-exc}
    M_{\bm k \bm k'} = \mathrm i \Xi \left(\frac{\hbar|\bm{k}-\bm{k}'|}{2\rho S s}\right)^{1/2}\cos{\varphi}.
\end{equation}
We are interested in excitons with an energy on the order of temperature. Since the main role in their scattering is played by phonons with the same momentum, the phonon energy is much less than $k_B T$, thus $N(q)$ takes the form:
\begin{equation}
\label{ph:stat}
    N(|\bm k-\bm k'|) = \frac{1}{\exp\left(\frac{\hbar s |\bm k - \bm k'|}{k_B T}\right)-1}\approx \frac{k_BT}{\hbar s|\bm k-\bm k'|}.
\end{equation}
After substituting~\eqref{matr:el:ph-exc},~\eqref{ph:stat} into~\eqref{tau:FGRSP}, we obtain the following expression for the energy-dependent exciton momentum relaxation rate
\begin{equation}
\label{tau:ph:fin}
    \frac{1}{\tau_{ph}(\varepsilon)} = \varepsilon\frac{k_BT\Xi^2}{2\rho\hbar^3s^2c^2_x}.
\end{equation}
Correspondingly, the diffusion coefficient of excitons at the scattering by long wavelength acoustic phonons reads
\begin{equation}
\label{Diffus:ph}
    D_{cl}^{ph}  = \frac{2}{(k_BT)^2}\frac{\rho\hbar^3s^2c_x^4}{2\Xi^2} \propto T^{-2}.
\end{equation}
As in the case of scattering by static defects, Eq.~\eqref{Diffus:def}, the phonon-induced diffusion coefficient decreases with increasing the temperature. The $D_{cl}^{ph} \propto T^{-2}$ dependence results from both combination of the linear-in-the-energy density of states making $\tau_{ph} \propto \varepsilon^{-1}$ and the fact that the number of available phonons increases $\propto T$, see Eq.~\eqref{ph:stat}.

\begin{figure}[t]
    \centering
    \includegraphics[width=\linewidth]{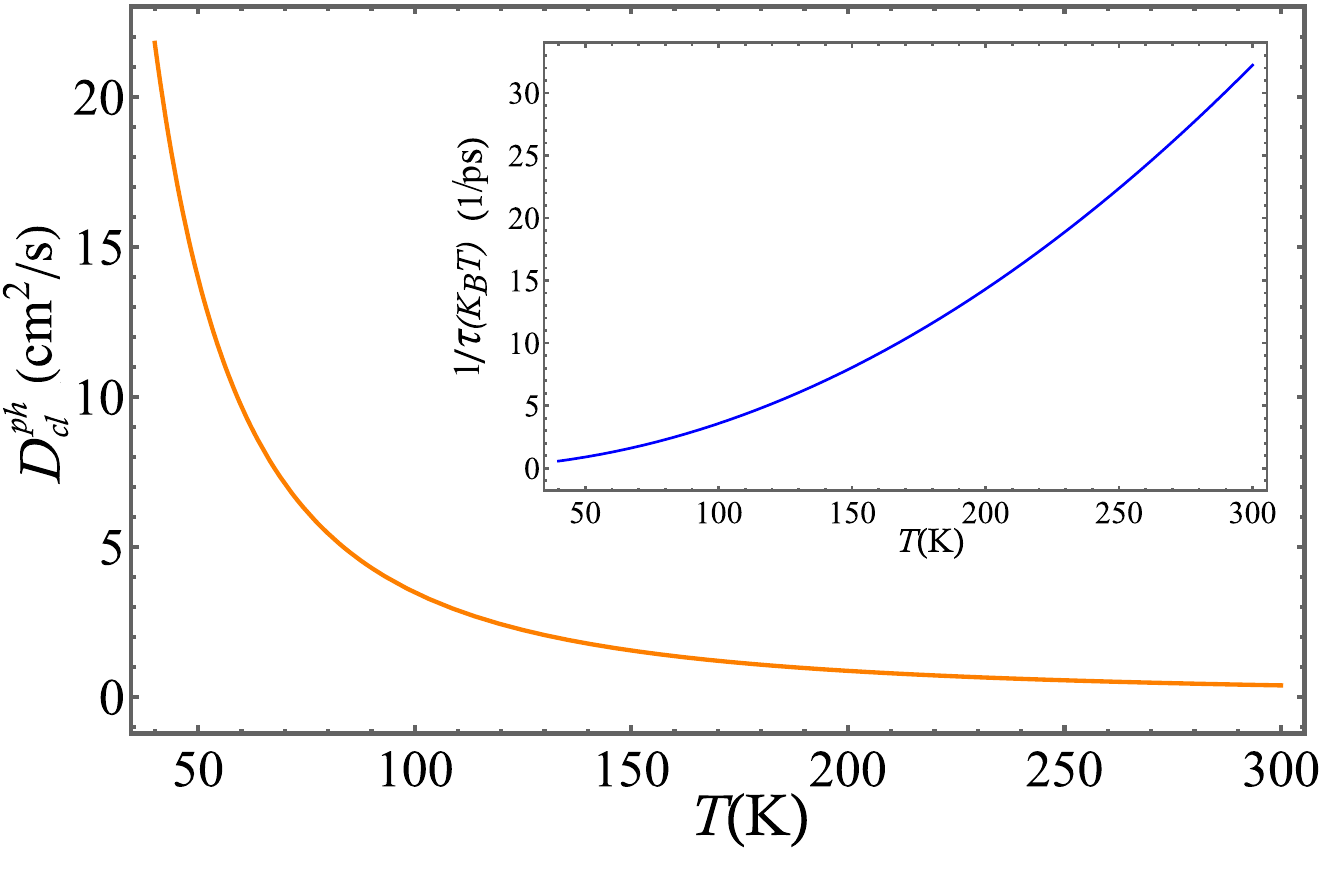}
    \caption{The temperature dependence of the diffusion coefficient for scattering by longitudinal acoustic phonons. Inset shows temperature dependence of the momentum relaxation rate, $1/\tau(\varepsilon)$, at $\varepsilon = K_BT$. The parameters of calculation are~\cite{shree2018observation}: $\rho = 4.5 \cdot 10^{-7}$ g/cm$^2$, s = $4.6 \cdot 10^{5}$ cm/s, $\Xi = 6.4$ eV, and $c_x = 5 \cdot 10^6$ cm/s in Eq.~\eqref{Diffus:ph}, see Sec.~\ref{subsec:est}.}
    \label{fig2}
\end{figure}

The dependence of $D_{cl}^{ph}$ is plotted in Fig.~\ref{fig2} for generic parameters relevant for transition metal dichalcogenide based moir\'e superlattices. The diffusion coefficient drops from $\sim 10$~cm$^2$/s to $0.1\ldots 1$~cm$^2$/s with the temperature decrease from $50$ to $300$~K {(at the temperatures close to the room one the scattering by the optical phonons and dispersion-less acoustic phonons becomes important).} The total diffusion coefficient with allowance for both phonon and impurity scattering can be obtained by substituting $1/\tau = 1/\tau_{ph} + 1/\tau_{def}$ into general Eq.~\eqref{eq:D:tau} with the result
\begin{equation}
    D_{cl} = \frac{1}{k_BT(1 + T/T_{d-p})}\frac{\hbar^3c_x^4}{nu^2},
\end{equation}
where
\begin{equation}
    T_{d-p} = \frac{\rho s^2nu^2}{k_B\Xi^2}.
\end{equation}
Naturally, as the temperature of the order of $T_{d-p}$, phonons start to control the diffusion coefficient. The temperature $T_{d-p}$ strongly depends on the density and type of static defects, i.e., on the properties of disorder that can significantly vary from sample to sample.

\subsection{\label{sec:hopping}Dipole-dipole interaction assisted hopping}
At sufficiently low temperatures, the semiclassical propagation scenario becomes irrelevant. It is because in the presence of arbitrary weak disorder the exciton states turn out to be localized despite the long-range $1/|\bm r_i - \bm r_j|^3$ decay of the transfer matrix elements between the sites $i$ and $j$ enabled by the electromagnetic field. Indeed, qualitatively, the localization-delocalization transition in a non-interacting system is controlled by possibility of forming an infinite cluster of resonant states whose energy splitting $\Delta E_{ij}$ is smaller than the coupling matrix element $M_{ij}$. Since in two-dimensional systems the number of available sites within a distance $R$ from a given one increases as $R^2$ and the matrix element $M_{ij} \propto R^{-3}$ the cluster cannot be formed~\cite{ES:book,Levitov:1989aa,footnote:crit}.

Localization of single-particle states implies that the hopping regime of transport at finite (but low enough) temperatures can be realized. We consider a variable range hopping scenario for excitons and only provide basic estimates for the diffusion coefficient. The development of the detailed theory of the hopping transport in such a case is beyond the scope of the present paper. The typical hopping distance is determined by interplay of (i) probability of finding a sufficiently close-in-energy localization site in the vicinity of a given one and (ii) the overlap integral between the wavefunctions of the localized states. Hence, the probability of a jump by the distance $R$ is given by:
\begin{equation}
    \label{mott:P}
    P(R) \propto \exp\left(-\frac{1}{\pi R^2 g k_B T} \right) \left(\frac{a_M}{R}\right)^6.
\end{equation}
Here the first, exponential, factor is responsible for the probability of finding a state in the area $\pi R^2$ in the energy band $k_B T$ with $g$ being the density of states (the number of states per unit energy and area, we assume it to be constant), and the second, power-law, factor is responsible for the overlap of wave functions at different localization sites. Naturally, longer-distance jumps are beneficial from an energy point of view: the larger $R$, the more likely it is to find a near-resonant site. However, the probability of such a transfer $\propto |M_{ij}|^2\propto R^{-6}$ decreases with increasing the distance $R$.

To provide an estimate for the diffusion coefficient we follow the approach presented in Ref.~\cite{ES:book} and, first, calculate a ``conductivity'' $\sigma$ of excitons in response to an artificial field $F$ assuming that they have a fake charge $e^*$ and, second, use the Einstein relation (for non-degenerate particles)
\begin{equation}
\label{Einst}
D = \frac{\sigma}{(e^*)^2 n} k_B T,
\end{equation}
to evaluate the diffusion coefficient. Using Eq.~\eqref{mott:P} the resistance corresponding a hop between the sites $i$ and $j$ separated by a distance $R_{ij}$ can be written as~\cite{ES:book}
\begin{equation}
\label{res:ij}
\mathcal R_{ij} = \frac{k_B T}{(e^*)^2 w} \frac{1}{\exp\left(-\cfrac{1}{\pi R_{ij}^2 g k_B T} - 6\ln{\frac{R_{ij}}{a_M}}\right)},
\end{equation}
where $w$  is the microscopic rate of exciton-phonon interaction. The latter is proportional to a power of the temperature. The critical resistance $\mathcal R_c$ corresponds to hops by the optimal distance $R_{opt}$ where two terms in the exponent controlling $\mathcal R_{ij}$ are approximately equal. At $R_{ij} \sim R_{opt}$ the exponential temperature dependence in~\eqref{res:ij} vanishes. According to the percolation theory, it determines the conductivity of a two-dimensional disordered system: $\sigma \propto \mathcal R_c^{-1}$. Hence, we conclude that the diffusion coefficient in the hopping regime 
\begin{equation}
    \label{D:hop:1}
    D_{hop} \propto T^\eta, \quad \eta>0,
\end{equation}
depends on the temperature in a power-law manner with the positive exponent $\eta$. It makes our system with electromagnetic-field (dipole-dipole interaction) assisted hopping quite different from conventional systems with localized excitations where the hopping regime is characterized by exponential dependence of the diffusion coefficient on the temperature~\cite{ES:book}. Simple estimate with $w\propto {\rm const}$ gives $\eta = 3$, however, the approach presented above does not allow us to determine $\eta$ precisely; this is a task requiring separate research beyond the scope of the present paper.

We also note that for $R_{opt}/a_M$ not too large the logarithm in the exponent in Eq.~\eqref{res:ij} can be approximated as $\ln{(R_{opt}/a_M)} \approx (R_{opt}/a_M)^{1/3}$~\cite{PhysRevB.105.054206,Huang:2022aa} that results in approximate exponent-like behavior with $D\propto \exp{[- 7(T_0/T)^{1/7}]}$ where $T_0 = (\pi g a_M^2)^{-1}$.

The analysis above is valid provided that the exciton recombination time $\tau_r$ is sufficiently long to let exciton pass the critical link (resistance). Otherwise, significant spatial fluctuations of the diffusion coefficient are expected.

\subsection{Discussion}

The analysis above demonstrates that electromagnetic field assisted exciton diffusion in moir\'e superlattices can be quite efficient. For instance, in the hopping regime, the power law dependence of the diffusion coefficient on the temperature is expected to provide faster exciton propagation as compared to the conventional case of exponential temperature dependence of the diffusivity, see, e.g.,~\cite{Akmaev:2021up,li2021interlayer}. In the semiclassical regime calculations presented in Fig.~\ref{fig2} show that the diffusion coefficient $D_{cl}$ is in the range of units to tens of cm$^2$/s that is comparable to or even larger than the diffusion coefficient of free excitons in monolayers and reconstructed, moir\'e-free bilayers~\cite{Chernikov:2023ab,PhysRevLett.132.016202,zipfel2019exciton}.

It is noteworthy that in addition to propagating longitudinal excitons there is an almost dispersionless branch of transverse excitons, see Eq.~\eqref{dispersion:T} and Fig.~\ref{fig1}(c,d). The energies of transverse excitons are lower than those of longitudinal ones, hence, one expects the longitudinal exciton lifetime to be shorter than the lifetime of transverse excitons due to (i) phonon-assisted scattering to the light cone states with $k\leqslant q$ and subsequent radiative decay and (ii) relaxation towards the lower-lying transverse excitons. These scattering processes limit the lifetimes of longitudinal excitons.

As a result, the propagation of a two-component system composed of fast moving and short lived longitudinal excitons and slow moving and longer lived transversal excitons can be more complex. For example, it can demonstrate effective negative diffusion where rapidly expanding longitudinal excitons recombine and give place to almost non-propagating transversal excitons. As a result the area of the spot occupied by excitons shrinks with time as if the diffusion coefficient were negative, $D<0$~\cite{PhysRevLett.132.016202,Rosati:2020aa,Ziegler2020,PhysRevB.107.045420,Kurilovich2024}. Analysis of such and other possible propagation scenarios~\cite{glazov2024ultrafastexcitontransportvan,mantsevich2024viscoushydrodynamicsexcitonsvan,adejumobi2024diffusionfastslowexcitons} including the processes of relaxation and decay of excitons is an interesting task for future. 

{We note that efficient energy relaxation towards low-energy transverse branch and low values of electromagnetic-field assisted exciton propagation velocity $c_x$, Eq.~\eqref{cx}, for spatially indirect excitons MoSe$_2$/WSe$_2$ heterostructure can be a possible cause of exciton diffusion with $D$ ranging from $10^{-2}$ to $10^{-1}$~cm$^2$/s recently observed in Ref.~\cite{li2021interlayer}. Hence, among various van der Waals heterostructures based on transition metal dichalcogenides where the exciton transport is studied~\cite{Tagarelli:2023aa,PhysRevLett.132.016202,li2021interlayer,upadhyay2024giantenhancementexcitondiffusion,yan2024anomalouslyenhanceddiffusivitymoire} probably MoSe$_2$/WS$_2$ heterobilayers are the most promising ones as they demonstrate type-I band alignment allowing for the high oscillator strength excitons and also the band aligment tunability~\cite{Kistner-Morris:2024aa}.}

\section{\label{sec:conclusion}Conclusion}

To conclude, we have demonstrated that dipole-dipole interactions, i.e.,  the coupling of localized excitons in moir\'e superlattices with induced electromagnetic field results in linear dispersion of longitudinal excitons whose microscopic dipole moment is parallel to the wavevector. It gives rise to propagation of excitons with a constant group velocity being $10^{-4}\ldots 10^{-3}$ of the speed of light depending on the system and type of excitons. Interaction with phonons and static disorder results in diffusive propagation of excitons with the semiclassical diffusion coefficient $\propto T^{-2}$ (at the long-wavelength acoustic phonon scattering) and $\propto T^{-1}$ (at the short-range defects scattering).

We have also briefly analyzed the localization of the excitons in the presence of disorder. Despite the long-range $\propto 1/r^3$ coupling between the lattice sites the single-particle states are localized by a disorder owing to the two-dimensional character of the system. The phonon-assisted variable-range hopping turns out to be quite unusual with the power law dependence of the diffusion coefficient on the temperature by contrast to conventional systems where the diffusion is described by the Mott's or Efros-Shklovskii's laws.

\acknowledgments
The authors are grateful to Ata\c{c} Imamo\v{g}lu and Boris Shklovskii for valuable discussions. This work was partially supported by the RSF project No. 23-12-00142 (analytical model by M.M.G.). A.M. Shentsev acknowledges support (numerical calculations) of the Ministry of Science and Higher Education of the Russian Federation (project no. FFWR-2024-0017).

\appendix
{\section{Derivation of the main equations}\label{sec:append}}

{Effective Hamiltonian~\eqref{Ham:0} and equations of motion~\eqref{p:latt} can be derived without assumptions that excitons are the point dipoles. To that end we following Ref.~\cite{ivchenko05a}, first, find the excitonic wavefunctions and energies at a lattice site and, second, take into account the inter site coupling via the electromagnetic field.}

{To solve the quantum mechanical problem for the excitonic states in isolated minimum $V_0^{e,h}(\bm r_{e,h})$ of the moir\'e lattice potential acting on the electron $e$ and hole $h$ with $\bm r_{e,h}$ being the position vector of the corresponding charge carrier:
\[
V^{e,h}(\bm r_{e,h}) = \sum_i V_0^{e,h}(\bm r_{e,h} - \bm r_i),
\]
we present the two-particle Schr\"odinger equation within the effective mass approach as 
\begin{multline}
\label{exciton:schr}
-\left(\frac{\hbar^2}{2m_e}\Delta_e + \frac{\hbar^2}{2m_h}\Delta_h\right)\psi_n(\bm r_e, \bm r_h) \\
+ \left[V_0^e(\bm r_e) + V_0^h(\bm r_h) + U(\bm r_e - \bm r_h)\right]\psi_n(\bm r_e, \bm r_h) \\
= (\hbar\omega_n - E_g) \psi_n(\bm r_e, \bm r_h).
\end{multline}
Here the subscript $n$ enumerates the states within the potential minumum, $\psi_n(\bm r_e, \bm r_h)$ and $\hbar\omega_n$ are their envelope functions and eigenenergies (note that the energies include the band gap $E_g$), $m_{e,h}$ are the charge carrier effective masses, and $U(\bm r)$ is the electron-hole attraction potential usually taken in the Rytova-Keldysh form~\cite{RevModPhys.90.021001}.}

 {The values of the exciton energy $\hbar\omega_n$ and particular shapes of the wavefunctions are determined by the interplay of the Coulomb attraction and size-quantization, see Refs.~\cite{ivchenko05a,Semina_2022}. We abstain from detailed analysis of the envelope function in the general case, but we have analyzed two simple limiting cases of (i) ``strong'' and (ii) ``weak'' exciton confinement. In the ``strong'' confinement regime the electron-hole attraction is relatively weak as compared to the size quantization energies of individual charge carriers. Hence, the two-particle wavefunction can be represented as a product of the electron and hole states:
\begin{subequations}
    \label{two-cases}
    \begin{equation}
        \label{strong}
        \psi_n(\bm r_e, \bm r_h) = \psi_e(\bm r_e) \psi_h(\bm r_h),
    \end{equation}
    while the energy $\hbar\omega_n$ is a sum of the band gap, electron and hole size-quantization energies with the small negative correction related to the Coulomb interaction. The functions $\psi_e(\bm r_e)$ and $\psi_h(\bm r_e)$ are determined from the single-particle Schr\"odinger equations. The weak confinement regime is valid provided that the typical scale of exciton confinement (i.e., typical size of the potential minumum or corresponding ``oscillator length'' for approximately parabolic $V_0^{e,h}(\bm r)$) $a_{loc} \ll a_{exc}$ where $a_{exc}$ is the two-dimensional exciton Bohr radius.
In a weak confinement regime the exciton is localized in a potential minimum as a whole, namely,
    \begin{equation}
        \label{weak}
        \psi_n(\bm r_e, \bm r_h) = \Psi(\bm R) \varphi(\bm \rho),
    \end{equation}
where $\Psi(\bm R)$ with $\bm R=(m_e \bm r_e + m_h \bm r_h)/(m_e+m_h)$ being the exciton center of mass coordinate and ${\varphi}(\bm r)$ with $\bm \rho = \bm r_e - \bm r_h$ being the relative coordinate. In this case the exciton energy is a difference of the band gap and two-dimensional exciton binding energy with a  correction related to the center of mass confinement. The weak confinement approximation is valid for $a_{loc} \gg a_{exc}$. For the moir\'e period of $a_M\sim 10$~nm and barrier heights on the order of $100$~meV the localization length $a_{loc}$ and $a_{exc}$ are comparable, both $\sim 1\ldots 3$~nm, and intermediate confinement regime can be realized where none of approximations, Eqs.~\eqref{strong} and \eqref{weak} is, strictly speaking, valid. 
\end{subequations}}

{Once the wavefunctions and energies are found, we represent the state of the crystal as~\cite{ivchenko05a}
\begin{equation}
    \label{state}
    |\Psi\rangle = |0\rangle + \sum_{n,i,\alpha} c_{n,i}^\alpha(t) |{n,i,\alpha}\rangle,
\end{equation}
where $|0\rangle$ is the ground state of the crystal where the valence band is filled and conduction bands are empty, $|{n,i,\alpha}\rangle$ is the state with the exciton $\psi_n(\bm r_e - \bm r_i, \bm r_h - \bm r_i)$ (at lattice site $i$) active in the linear polarization $\alpha=x$ or $y$ being excited, and $c_{n,i}^\alpha(t)$ are the decomposition coefficients. This expression is valid provided that the exciton overlap between the neighbouring sites is weak $$\left|\int d\bm r_e d\bm r_h \psi_n^*(\bm r_e - \bm r_i, \bm r_h - \bm r_i)\psi_{n'}(\bm r_e - \bm r_j, \bm r_h - \bm r_j)\right| \ll 1$$ ($n\ne n'$, $i\ne j$). It corresponds to a tight-binding approximation for excitons~\cite{Suris:2015aa}. Note that this approximation is valid even if the distance between the sites is smaller than the wavelength of radiation induced by the excitons. The dielectric polarization induced by the exciton is given by
\begin{equation}
    P_{n,i}^\alpha(\bm r, t) = c_{n,i}^\alpha(t)  d_{cv}\psi_{{n}}(\bm r{ - \bm r_i}, \bm r{ - \bm r_i}) + {\rm c.c.},
\end{equation}
where $d_{cv}$ is the interband dipole matrix element and the envelope function of the in-plane motion of electron and hole taken at the coinciding coordinates of the charge carriers. Taking into account that the interaction of excitons with electromagnetic field is described as
\[
-\int d\bm r \bm P_{n,i}(\bm r,t) \bm E(\bm r, t),
\]
and expressing the electric field via the Greens function~\eqref{Greens} [cf. Eq.~\eqref{field:greens}] we obtain from the time-dependent Schr\"odinger equation for the function $\Psi$ a set of equations for the coefficients $c_{n,i}^\alpha(t) = c_{n,i}^\alpha \exp{(-\mathrm i \omega t)}$ in the form
\begin{equation}
    \label{set:c:gen}
    \hbar(\omega_{n}-\omega) c_{n,i}^\alpha  = 4\pi q^2 \sum_{n', j,
    \beta} |d_{cv}|^2 G_{\alpha,\beta}^{n,n'}(i,j) c_{n',j}^\beta, 
\end{equation}
where the summation is carried out over all lattice sites $j$ and all states $n'$ with the exclusion of the self-interaction term ($n=n'$, $j=j'$)
\begin{multline}
    \label{Gij}
    G^{n,n'}_{\alpha\beta}(i,j) = \int d\bm r d\bm r'\psi_n^*(\bm r- \bm r_i,\bm r - \bm r_{i})\\
    G_{\alpha,\beta}(\bm r -\bm r') \psi_{n'}(\bm r' - \bm r_j, \bm r' - \bm r_j).
\end{multline}
The set of Eqs.~\eqref{set:c:gen} is general and accounts for all excitonic states. It can be readily generalized to allow for the quantum mechanical tunneling of excitons: In that case the right hand side acquires additional term in the form $\sum_{n',j,\beta} \delta_{\alpha\beta} T^{n,n'}(i,j) c^\beta_{n',j}$ with $T^{n,n'}(i,j)$ being the transfer integrals~\cite{Suris:2015aa} and the Kronecker $\delta$-symbol ensures the conservation of polarization in the course of the tunneling.}

{Taking into account that typical spatial scale in the Greens function~\eqref{Greens} is the light wavelength corresponding to the exciton transition frequency ($2\pi/q \sim 10^2$~nm \ldots $10^3$~nm) is much larger than both the moir\'e lattice period $a_M \sim 10$~nm and both exciton Bohr radius and exciton localization length $a_{exc}$, $a_{loc}\sim 1\ldots 3$~nm we have 
\begin{multline}
    G^{n,n'}_{\alpha\beta}(i,j) = G_{\alpha\beta}(\bm r_i - \bm r_j)\\
    \int d\bm r d\bm r'\psi_n^*(\bm r,\bm r )
    \psi_{n'}(\bm r', \bm r' ), \quad i \ne j.
\end{multline}
If we disregard the coupling between the different orbital states of the excitons $n\ne n'$, the set of Eqs.~\eqref{set:c:gen} is equivalent to Eqs.~\eqref{p:latt} where $n=0$ is taken and, in agreement with Refs.~\cite{ivchenko05a,PhysRevA.109.053523} and Eq.~\eqref{d:exc:est}
\begin{equation}
    \label{d:exc:0}
    d_{exc} = d_{cv} \int d\bm r\psi_0(\bm r, \bm r).
\end{equation}
Note that in the strong confinement regime [Eq.~\eqref{strong}] $d_{exc} \approx d_{cv}$. In the weak confinement regime [Eq.~\eqref{weak}] $d_{exc} \sim d_{cv} (a_{loc}/a_{exc})$.}

{It is possible to show that the interstate coupling is weak: It follows from Eq.~\eqref{Gij} that the corresponding coupling matrix element is, roughly, the radiative broadening of a single localized exciton state~\cite{ivchenko05a}. For the parameters of the studied system it is on the meV scale (and localization typically suppresses it)~\cite{PhysRevLett.123.067401} while the interstate energy distances are on the order of tens to hunderds meV.}

{If the moir\'e lattice period $a_M$ is comparable to the light wavelength $2\pi/q$ the Bragg diffraction of light on the moir\'e lattice may occur~\cite{ivchenko05a}. This regime is beyond the scope of the present work.}
\providecommand{\noopsort}[1]{}\providecommand{\singleletter}[1]{#1}%


\begin{thebibliography}{59}
\bibitem{note:tun} {Strong Coulomb attraction between an electron and a hole prevents independent tunneling of charge carriers. Hence, the tunneling is mainly controlled by the exciton translational motion mass being the sum of electron and hole effective masses.}
\bibitem{footnote:crit} {The localization in our case is a result of the fact that the electromagnetic field is three-dimensional while excitons can propagate in a two-dimensional plane only. If the photons were two-dimensional as well, e.g., if the moir\'e lattice is placed in a waveguide, the $M_{ij} \propto 1/R^2$ and the localization-delocalization properties of such system turn out to be very different~\cite{PhysRevLett.107.076401,PhysRevB.109.174208}, see also~\cite{Burin89,PhysRevLett.64.547}.}
\end{thebibliography}
\end{document}